# Enhancing Efficiency in Computational Intensive Domains via Redundant Residue Number Systems

Soudabeh Mousavi[1], Dara Rahmati[1], Saeid Gorgin[2], Jeong-A Lee[2]

[1] Department of Computer Science and Engineering, Shahid Beheshti University, Tehran, Iran
[2] Department of Computer Engineering, Chosun University, Gwangju, Republic of Korea
soudabeh.mousavi99@gmail.com, d_rahmati@sbu.ac.ir, gorgin@chosun.ac.kr, jalee@chosun.ac.kr

*Abstract*— In computation-intensive domains such as digital signal processing, encryption, and neural networks, the performance of arithmetic units, including adders and multipliers, is pivotal. Conventional numerical systems often fall short of meeting the efficiency requirements of these applications concerning area, time, and power consumption. Innovative approaches like residue number systems (RNS) and redundant number systems have been introduced to surmount this challenge, markedly elevating computational efficiency. This paper examines from multiple perspectives how the fusion of redundant number systems with RNS (termed R-RNS) can diminish latency and enhance circuit implementation, yielding substantial benefits in practical scenarios. We conduct a comparative analysis of four systems—RNS, redundant number system, Binary Number System (BNS), and Signed-Digit Redundant Residue Number System (SD-RNS)—and appraise SD-RNS through an advanced Deep Neural Network (DNN) utilizing the CIFAR-10 dataset. Our findings are encouraging, demonstrating that SD-RNS attains computational speedups of 1.27 times and 2.25 times over RNS and BNS, respectively, and reduces energy consumption by 60% compared to BNS during sequential addition and multiplication tasks.

## I. Introduction

In recent years, technological advancements have led to an exponential increase in data generation, outpacing the computational capabilities of current systems. On the other hand, the efficiency of compute-intensive applications—such as neural networks [1], digital signal processing, and image processing—heavily depend on the speed of computer arithmetic units like adders and multipliers. However, in conventional binary number systems (BNS), computation speed diminishes as the number of bits increases, primarily due to carry propagation. Consequently, innovative approaches utilizing unconventional number systems, such as the residue number system (RNS), redundant number system, and redundant residue number system (R-RNS), have been explored to boost performance and efficiency in computations. RNS has emerged as a significant topic in fast computer arithmetic. In this numerical system, multiplication and addition operations are decomposed into smaller, independent units with fewer bits, thereby reducing carry digit propagation compared to BNS—though carry propagation within each digit persists [2]. Furthermore, redundant number systems are instrumental in executing rapid computations. Their defining characteristic is eliminating carry propagation between digits during addition/subtraction operations. When integrated with residue arithmetic—especially in applications where core tasks hinge on addition and multiplication—this attribute leads to more efficient computer arithmetic circuit implementations [4]. The advantages of R-RNS have been documented in several studies [3]. Similarly, SD-RNS (Signed-Digit Residue Number System) is noted for its speed and efficiency across various computational paradigms [4]. In this paper, we evaluate the speed of adders and multipliers within these four numerical systems: BNS, RNS, redundant number representation, and SD-RNS. We then propose a framework for selecting the most appropriate numerical system for diverse applications tailored to their specific computational needs.

## II. System Overview

This study compares the efficiency of addition and multiplication in hardware implementations, focusing on applications that require frequent arithmetic operations within a defined numerical range. It assesses four computational methods: BNS, RNS, Redundant Number Representation, and SD-RNS. An $n-$digit SD integer, $X = [x_{n-1}, \ldots, 0]$, $x_i \in \{\bar{1}, 0, 1\}$ has the value

$$X = \sum_{i=0}^{n-1} x_i 2^i \qquad (1)$$

This representation offers multiple ways to represent an integer, making it a redundant number system. In this work, multiplier and adder circuits have been implemented in different systems using a $\{2^n - 1, 2^n, 2^n + 1\}$ set of modules besides the BNS at various precisions. The adder and multiplier cells of SD and SD-RNS used in this project are circuits described in [3]. An SD adder consists of identical adder cells operating in parallel and independently of each other, while an SD-RNS adder requires a carry end-around operation. The main difference between SD and SD-RNS multiplier circuits is the number of input bits. In RNS, operands are first converted into remainders according to the modules of the system, which are smaller numbers. Arithmetic operations are performed on each remainder after redundant representation is applied to each remainder. The partial products $pp_i = \langle 2^{2i} \cdot \langle rp_i \rangle \rangle_m$ can be generated by rotation if the moduli are $2^n$, $2^n - 1$ or $2^n + 1$ by using the rules in equation set Eq. 2. These rotations can be accomplished by wiring connections appropriately [3].

$$\langle 2^a \cdot y \rangle_{2^p-1} = [y_{p-1-a} \ldots y_0 y_{p-1} \ldots y_{p-a}]$$
$$\langle 2^a \cdot y \rangle_{2^p} = [y_{p-1-a} \ldots y_0 0_{p-1} \ldots 0_{p-a}]$$
$$\langle 2^a \cdot y \rangle_{2^p+1} = [y_{p-1-a} \ldots y_0 (-y_{p-1}) \ldots (-y_{p-a})]$$
$$p \geq a, y = [y_{p-1} \ldots y_0] \qquad (2)$$

The RNS adder and multiplier for the selected set of modules comprise three adders and multipliers in channels $2^n, 2^n - 1$, and $2^n + 1$. The adder is implemented using the proposed architecture in [4] based on the Sklansky prefix structure. For the hardware implementation of the multiplier, for all the channels, the radix-4 Booth encoding (as a partial product generator) is utilized. These circuits have been implemented using the proposed architecture described in [5]. The overall delay of the four mentioned systems for an operation involving a forward conversion, a reverse conversion, $x$ addition operations, and $y$ multiplication operations are as follows:

$$T_{Total} = \{T_{FC}\} + x\{T_{Add}\} + y\{T_{Mul}\} + \{T_{RC}\} \qquad (3)$$

## III. EVALUATION

The performance assessment of the four numerical systems was conducted using AlexNet and VGG16 networks alongside the CIFAR-10 dataset. Results were analyzed for precisions of 16, 24, 32, and 64 bits, with channel sizes of $n = 5, 8, 11$, and 21 in the moduli set $\{2^n - 1, 2^n, 2^n + 1\}$, as depicted in Table I. Fig. 1 presents delay plots for the numerical systems, taking into account three parameters: $x$ (number of additions), $y$ (number of multiplications), and $z$ (delay due to multiplication and addition operations). The plots indicate that SD-RNS excels in applications with balanced numbers of consecutive additions and multiplications and in programs dominated by consecutive multiplications, owing to its rapid computation capabilities. Conversely, the redundant residue number system demonstrates superior speed for tasks focused exclusively on addition/subtraction operations, achieving addition in constant time irrespective of operand length. The analysis reveals that in all scenarios, the delay of SD-RNS is consistently lower than that of the RNS system outlined in Table II. This table demonstrates that for various applications, which may require anywhere from no to many additions followed by multiplications, the most suitable number representation system—RNS, SD-RNS (or R-RNS), and SD—can be deduced from the table.

Table I.
Synthesis results with module set $\{2^n - 1, 2^n, 2^n + 1\}$ in four systems

| Circuit | Delay(ns) | | | |
| --- | --- | --- | --- | --- |
| | $P = 16$, $n = 5$ | $P = 24$, $n = 8$ | $P = 32$, $n = 11$ | $P = 64$, $n = 21$ |
| SD module adder | 0.21 | 0.21 | 0.21 | 0.21 |
| RNS module adder | 0.28 | 0.37 | 0.42 | 0.58 |
| SD adder | 0.21 | 0.21 | 0.21 | 0.21 |
| BNS adder | 0.3 | 0.38 | 0.45 | 0.63 |
| SD module multiplier | 0.43 | 0.63 | 0.74 | 0.97 |
| RNS module multiplier | 0.5 | 0.72 | 0.84 | 1.28 |
| SD multiplier | 0.8 | 0.98 | 1.03 | 1.24 |
| multiplier | 1.05 | 1.28 | 1.5 | 1.9 |

Table II.
The results of selecting different numerical systems with varying numbers of multiplications and additions

| Adder \ Multiplier | Zero | Low | Medium | High |
| --- | --- | --- | --- | --- |
| Zero | - | SD-RNS/RNS | SD-RNS/RNS | SD-RNS |
| Low | SD | SD-RNS/RNS | SD-RNS/RNS | SD-RNS |
| Medium | SD | SD-RNS | SD-RNS/RNS | SD-RNS |
| High | SD | SD-RNS | SD-RNS | SD-RNS |

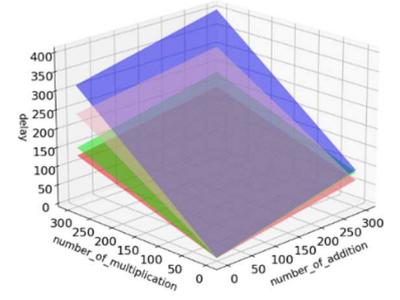

$P = 16, n = 5$

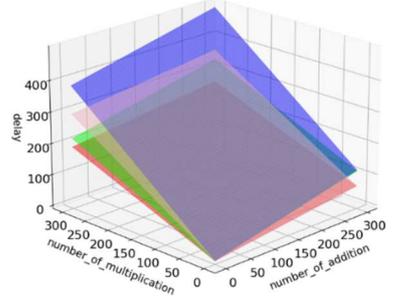

$P = 24, n = 8$

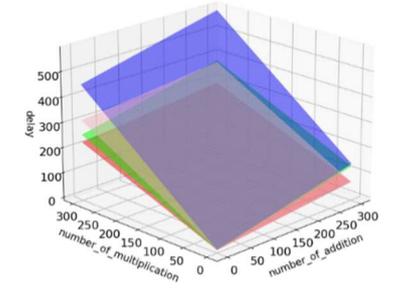

$P = 32, n = 11$

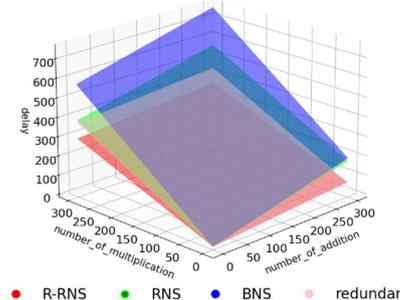

$P = 64, n = 21$

● R-RNS ● RNS ● BNS ● redundant

Fig.1 The delays of the four systems with the number of multiplication and addition operations with module set $\{2^n - 1, 2^n, 2^n + 1\}$

## IV. CONCLUSION

In this research, we conducted a comparative study of four number representation systems and developed an SD-RNS-based architecture to improve the efficiency of DNN computational circuits, followed by a thorough performance evaluation. The implementation of SD-RNS in applications that rely solely on arithmetic operations, such as addition and multiplication—encompassing digital signal processing systems, digital filters, encryption algorithms, and neural networks—results in a notable increase in computational speed and energy efficiency. The efficacy of the proposed architecture was assessed using the AlexNet and VGG-16 networks with the CIFAR-10 dataset. The findings reveal that SD-RNS achieves a 1.27-fold increase in calculation speed over RNS and a 2.25-fold enhancement over BNS. Additionally, this paper determines the most appropriate number representation system—RNS, and SD-RNS—based on the specific requirements of a wide range of applications, from those with minimal to extensive addition and multiplication operations.